\documentclass[preprint2]{aastex}

\usepackage{amssymb,natbib}
\defcitealias{Stacey.Charmandaris.Boulanger.2010}{SCB10}

\shorttitle{Mid-J CO Emission From NGC 891
}
\shortauthors{Nikola et al.}

\begin{document}

\title{Mid-J CO Emission From NGC 891: Microturbulent Molecular Shocks in Normal Star Forming Galaxies}

\author{T. Nikola\altaffilmark{1}, G. J. Stacey\altaffilmark{1}, D. Brisbin\altaffilmark{1}, C. Ferkinhoff\altaffilmark{1}, S. Hailey-Dunsheath\altaffilmark{2}, S. Parshley\altaffilmark{1}, C. Tucker\altaffilmark{3}}

\altaffiltext{1}{Cornell University, Ithaca, NY 14853}
\altaffiltext{2}{Max-Planck-Institut f\"ur extraterrestrische Physik, Germany}
\altaffiltext{3}{Cardiff University, UK}

\begin{abstract}
We have detected the CO(6-5), CO(7-6), and [C~{\small I}] 370~$\mu$m lines from the nuclear region of NGC~891 with our submillimeter grating spectrometer ZEUS on the CSO.
These lines provide constraints on photodissociation region (PDR) and shock models that have been invoked to explain the H$_{2}$~S(0), S(1), and S(2) lines observed with Spitzer.
We analyze our data together with the H$_{2}$ lines, CO(3-2), and IR continuum from the literature using a combined PDR/shock model.
We find that the mid-J CO originates almost entirely from shock-excited warm molecular gas; contributions from PDRs are negligible. 
Also, almost all the H$_{2}$ S(2) and half of the S(1) line is predicted to emerge from shocks.
Shocks with a pre-shock density of  $2\times10^{4}$~cm$^{-3}$ and velocities of 10~km/s and 20~km/s for C-shocks and J-shocks, respectively, provide the best fit.
In contrast, the [C~{\small I}] line emission  arises exclusively from the PDR component, which is best parameterized by a density of $3.2\times10^{3}$~cm$^{-3}$ and a FUV field of $G_{o} = 100$ for both PDR/shock-type combinations.
Our mid-J CO observations show that turbulence is a very important heating source in molecular clouds, even in normal quiescent galaxies.
The most likely energy sources for the shocks are supernovae or outflows from YSOs.
The energetics of these shock sources favor C-shock excitation of the lines.
\end{abstract}

\keywords{turbulence --- galaxies: ISM --- galaxies: star formation --- galaxies: individual: NGC 891 --- submillimeter: ISM --- techniques: spectroscopy}

\section{Introduction \label{sec:intro}}

Turbulence and shocks play a profound role in the process of star formation and nuclear gas accretion.
Shocks compress the interstellar medium (ISM) and can trigger star formation, while
turbulence prevent the collapse of the ISM.
These mechanisms can also remove angular momentum from the gas, leading to gas infall toward galactic nuclei, a process that is important in interacting galaxies.

Both mechanisms are very important in converting mechanical energy into thermal energy and can dominate the heating budget of molecular clouds.
The main coolants of the shock heated molecular gas are H$_{2}$ and CO lines.
With their improved sensitivity, recent space observatories (ISO, Spitzer) were able to trace extragalactic rotational H$_{2}$ emission and revealed the importance of this emission and the prevalence of shock excited molecular gas.
For example, H$_{2}$ emission associated with shock or microturbulence excited molecular gas was found in the group Stephan's Quintet \citep{Appleton.Xu.Reach.2006, Cluver.Appleton.Boulanger.2010}, the radio galaxy 3C~326 \citep{Ogle.Antonucci.Appleton.2007}, nearby edge-on galaxies NGC 4565, NGC 5907 \citep{Laine.Appleton.Gottesman.2010} and NGC 891 \citep[hereafter SCB10]{Stacey.Charmandaris.Boulanger.2010}, as well as in the extended cold neutral medium in the Milky Way \citep{Falgarone.Verstaete.PineauDesForets.2005}.
Ground-base mid-J CO observations are also good probes of shock or turbulence excited molecular gas.
\citet{Bradford.Stacey.Nikola.2005} have invoked dissipation of MHD turbulence to explain their mid-J CO observations of the circumnuclear ring in Galactic center.
In the nuclear region of NGC 253, the strong mid-J $^{12}$CO \citep[e.g.][]{Bradford.Nikola.Stacey.2003} and $^{13}$CO(6-5) \citep{HaileyDunsheath.Nikola.Stacey.2008} emission is consistent with heating of the warm molecular gas by either decay of supersonic turbulence through shocks or by an enhanced cosmic ray density.

Due to its proximity (9.5 Mpc), and edge-on presentation \citep[$i\geq89^{\circ}$;][]{Sancisi.Allen.1979}, NGC 891 is the primary target for studies of the effects of energetic processes associated with star formation on the interstellar medium in normal spiral galaxies. 
The edge-on view of NGC~891 results in greatly enhanced gas columns so that relatively weak lines such as the low-lying quadrupole rotational transitions of H$_{2}$ become observable.  
The mid-IR H$_{2}$ rotational transitions in NGC~891 were studied for the first time with ISO \citep{Valentijn.vdWerf.1999} and then more recently with Spitzer \citepalias{Stacey.Charmandaris.Boulanger.2010}.
The latter study combined the H$_{2}$ lines observed with Spitzer (S(2), S(1), and S(0)) together with [C~{\small II}], [O~{\small I}] and IR continuum observations from ISO to constrain gas excitation mechanisms.  
They found that most of the S(0) line emission arises in photodissociation regions (PDRs), and most of the S(2) line emission arises in low velocity microturbulent shocks, with the S(1) line arising from both types of sources in the ratio 70\% PDRs, and 30\% shocks. 

Here, we expand on the analysis of \citetalias{Stacey.Charmandaris.Boulanger.2010} by including submillimeter observations of the CO(6-5), CO(7-6), and [C~{\small I}] 370~$\mu$m lines.
The mid-J CO lines can arise from gas heated by a variety of mechanisms, including FUV photons (PDR), X-rays (XDR), cosmic rays, and shocks. 
The [C~{\small I}] emission arises mainly from PDRs.
Since the Spitzer H$_{2}$ study showed that the main heating sources of the warm molecular gas in NGC~891 are turbulence and PDRs we will focus on these two heating mechanisms and use our new submillimeter lines to both test and constrain the models of \citetalias{Stacey.Charmandaris.Boulanger.2010}.
As in their analysis, we apply the PDR model of \citet{Kaufman.Wolfire.Hollenbach.1999} and the shock model of \citet{Flower.PineauDesForets.2010}.

\section{Observations \label{sec:obs}}

We have observed the $^{12}$CO $J=6\to5$ (433.56 $\mu$m) and $J=7\to6$ (371.65 $\mu$m) rotational transitions and the [C~{\small I}] $^3P_2 \to ^3P_1$ (370.41 $\mu$m) fine structure line with our submillimeter grating spectrometer ZEUS \citep{HaileyDunsheath.2009} on the Caltech Submillimeter Observatory (CSO) on Mauna Kea, Hawaii, in December 2006.

ZEUS employs a $1\times32$ thermistor sensed detector array that is optimized for observations in the submillimeter wavelength regime.
It provides 32 spectral elements at one spatial position. 
For the Dec.\ 2006 observing run, we mounted two bandpass filter appropriate for the 350 and 450 $\mu$m telluric window adjacent to each other at the entrance slit of the detector housing, each filter covering 16 spectral pixels.
This allowed us to observe in the 350 and 450 $\mu$m telluric windows simultaneously and it still provided a full bandwidth coverage of about 5000~km/s in each band.
The spectral resolution of ZEUS is $\lambda/\Delta\lambda \sim 1000$ across the 350 $\mu$m and 450 $\mu$m telluric windows. 
The CO(7-6) and [C~{\small I}] lines, which are separated in velocity by only 1001~km/s, were observed simultaneously within a single spectrum.
We verified the spectral calibration through observation of CO gas absorption lines from a calibration unit mounted in front of the ZEUS dewar.

The CO lines were observed on two different nights.
On the first night we observed only the CO(6-5) emission at the position RA=${\rm 02^h22^m33.10^s}$ and Dec=$+42^{\circ}20'55.1''$ (J2000).
On the second night we observed the CO(6-5), (7-6), and [C~{\small I}] lines, but centered $\sim 2''$ further to the north.
This difference is, however, within the pointing accuracy of the telescope.
All observations were made in chop-nod mode with the secondary chopping at 2~Hz and a chop throw of $30''$.

We established a pointing model from observations of Saturn and Uranus at various elevation and azimuth positions.
The derived pointing accuracy is $\lesssim 5''$. 
We estimated the coupling of the planets to the beam by assuming the planets to be disk-like sources with uniform emissivity and assuming the ZEUS beam to be Gaussian. 
During the observations the beam size was $10''$ in the 350 $\mu$m band and $11''$ in the 450 $\mu$m band. 
At the distance of NGC~891, $10''$ corresponds to 460~pc.

All observations were corrected for atmospheric absorption through use of the CSO radiometers.  
Each spectrum was corrected for the response function of the grating and detector array, then flux calibrated with respect to Saturn, which was assumed to emit like a blackbody with a temperature of $T = 113$~K at 370~$\mu$m and 434~$\mu$m \citep{Hildebrand.Loewenstein.Harper.1985}. 
The correction for telluric transmission is the largest source of error.
Overall, we estimate a line flux uncertainty of $\sim 25$\%.  

\begin{figure}[ht]
\begin{minipage}[t]{5cm}
\centering
\includegraphics[width=0.90\textwidth]{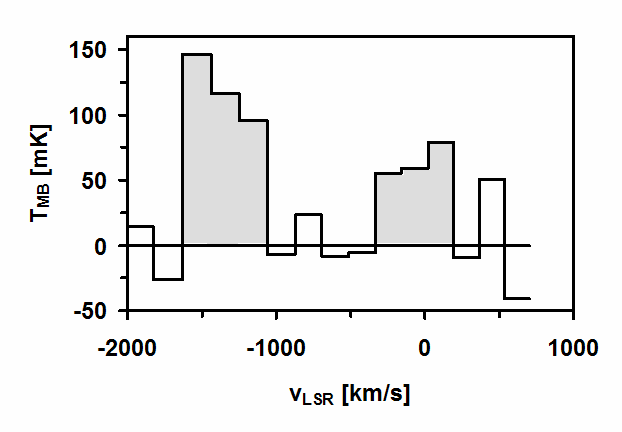}
\end{minipage}
\vspace{0.2cm}
\begin{minipage}[t]{5cm}
\centering
\includegraphics[width=0.90\textwidth]{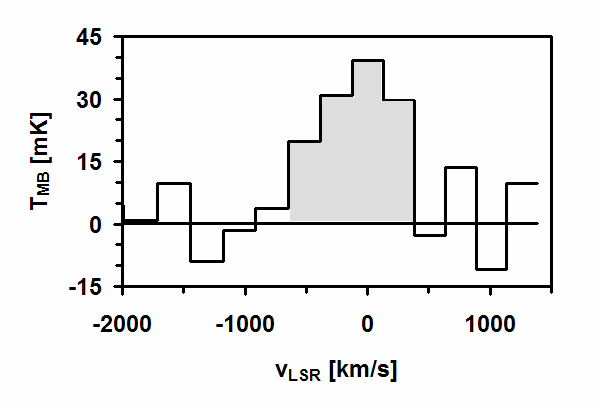}
\end{minipage}
\caption{ZEUS [C~{\small I}] and CO(7-6) (top), and CO(6-5) (bottom) spectra.}
\label{fig:spec}
\end{figure}

The CO(6-5), CO(7-6), and [C~{\small I}] 370 $\mu$m lines were all strongly detected (Fig.~\ref{fig:spec}), and their observed line intensities are listed in Table~\ref{tab:zeusobs}.

\section{Origin Of The Mid-J CO Emission in NGC 891 \label{sec:origin}}
 
The warm molecular gas traced by mid-J CO emission can be excited by several mechanisms.
For spiral galaxies with quiescent star formation, the most relevant are shocks in micro-turbulent gas or far-UV radiation in photodissociation regions (PDRs).
Spitzer observations of the H$_{2}$ mid-IR rotational transitions in NGC~891 have revealed the presence of warm, shock-excited molecular gas.
The line intensities can only be explained if part of the H$_{2}$ emission arises from this warm, micro-turbulent gas phase \citepalias{Stacey.Charmandaris.Boulanger.2010}.
While the H$_{2}$ S(0) likely originates entirely from PDRs, 30\% of the H$_{2}$ S(1) and the majority (80\%) of the H$_{2}$ (S2) emission likely emerges from gas excited by C-type shocks with a pre-shock gas density of $\sim 2 \times 10^{4}$ cm$^{-3}$, shock velocities between 20-30 km/s, and a shock filling factor of 2.8 \citepalias{Stacey.Charmandaris.Boulanger.2010}. 
The properties of this shock excited gas should also be imprinted in the line ratio of our CO observations.

We are taking a slightly different approach in our shock-analysis than \citetalias{Stacey.Charmandaris.Boulanger.2010}.
For their analysis they considered H$_{2}$ emission from within large areas of the galaxy since they compared their Spitzer observations with far-IR lines obtained with ISO, which had a beam size of $75''$.
Here, we focus on the central $11''$ of NGC~891 and with our new ZEUS mid-J CO observations we can refine the shock modeling.
We combine the Spitzer H$_{2}$ observations (Table~\ref{tab:litobs}) with our ZEUS observations and apply the shock models of \citet{Flower.PineauDesForets.2010}.
The spatial resolution of the Spitzer H$_{2}$ observations are very close to our ZEUS observations. 
The slit orientation of the long-high IRS Spitzer H$_{2}$ S(0) observation was such that the slit length ($22.3''$) is along the major axis and the slit width ($11.1''$) is perpendicular to the major axis.
For the short-high IRS Spitzer observations of the H$_{2}$ S(1) and S(2) lines, the slit is rotated, with the slit length ($11.3''$) oriented perpendicular to the major axis of NGC~891 and the slit width ($4.7''$) along the plane of the galaxy. 
Thus perpendicular to the plane of the galaxy the slits cover about the same area. 
This width is also nearly identical to the FWHM of the ZEUS beam. 
However, there is a gap of $30''$ between the individual positions of the Spitzer observations along the major axis. 
Hence, to convert the Spitzer H$_{2}$ observations to the ZEUS beam size we have to assume a distribution of the H$_{2}$ emission. 

In our analysis, we also include the CO(3-2) line, which is available in the literature \citep{Mauersberger.Henkel.Walsh.1999, Dumke.Nieten.Thuma.2001, Bayet.Gerin.Phillips.2006}.
Unfortunately, the line intensities and flux densities are not consistent.
Despite very similar beam sizes, the reported values vary by a factor of about 6.5.
For our analysis we use the intensity given by \citet{Mauersberger.Henkel.Walsh.1999} (Table~\ref{tab:litobs}), which are about a factor of three less than the largest reported intensity.

The CO(1-0) \citep[e.g.][]{Scoville.Thakkar.Carlstrom.1993, Sofue.Nakai.1993} and CO(3-2) \citep{Mauersberger.Henkel.Walsh.1999, Dumke.Nieten.Thuma.2001} emission shows a plateau of extended emission with strong central peak and local maxima roughly $1'$ to the North-East and South-West. 
The H$_{2}$ Spitzer observations also show extended emission with a central peak, a trough at the next sampling positions $30''$ away from the center, local maxima at the second sampling position at $60''$ from the center on either side, and a decrease in the following sampling positions. 
Thus, within the limited sampling of the Spitzer observations the distribution of the H$_{2}$ and CO emission appear similar. 
\citet{Dumke.Nieten.Thuma.2001} estimated a source size of about $17''\times8''$ (deconvolved) for the central gas concentration from their CO(3-2) map. 
The profile of the CO(1-0) emission along the major axis is similar, but slightly larger, showing a FHWM of about $25''$ above the plateau \citep[Fig.~2 and 3 in][, respectively]{Scoville.Thakkar.Carlstrom.1993, Sofue.Nakai.1993}. 
It is plausible that the higher excitation CO emission is more concentrated toward the center than the CO(1-0) emission. 
Here, we use the source size estimated from the CO(3-2) emission to convert the H$_{2}$ Spitzer observations and the CO(7-6) and [C~{\small I}] intensities to the $11''$ beam size of our ZEUS CO(6-5) observations, resulting in conversion factors of 1.4 for the H$_{2}$ S(0) line and 0.9 for the H$_{2}$ S(1) and S(2) lines. 
The converted intensities are given in Table~\ref{tab:fit_res}.

To model the shock parameters we use a different method than \citetalias{Stacey.Charmandaris.Boulanger.2010}.
They compared the H$_{2}$ emission with the far-infrared ISO lines on the scale of the large ISO beam, determined the PDR model that best fits the data (biased toward the ISO far-IR lines), and attributed the remaining H$_{2}$ emission to shock-excited, micro-turbulent gas.
Since the ZEUS observations cover a much smaller region we can not use the ISO observations to constrain the line emission fraction.
Scaling the ISO observations to the much smaller beam size of ZEUS is subject to large uncertainties without knowing the morphology of the gas-phase locked in PDRs.
While a reasonable scaling factor could be inferred for some far-IR lines, it would be questionable for others.
Several observations with the Kuiper Airborne Observatory (KAO) have found that the distribution of the [C~{\small II}] and CO(1-0) emission are correlated in galaxies \citep[cf.][]{Crawford.Genzel.Townes.1985, Stacey.Geis.Genzel.1991}.
It is therefore reasonable to use the observed CO(1-0) morphology in NGC~891 to scale the [C~{\small II}] ISO observation to the smaller ZEUS beam.
However, no such correlation has been shown for the [O~{\small I}] lines.
In our approach we therefore started by only using the higher spatial resolution observations, the CO(7-6), (6-5), and CO(3-2) transition together with the the H$_{2}$ S(0), S(1) and S(2) lines, and applied a constrained least squares algorithm to simultaneously fit the individual line emission fraction, beam filling factor, and the shock parameters for just a C-shock and a J-shock model.
Unfortunately, trying to fit all these parameter to just a shock model results in an under-constrained problem and no unique solution is possible.
This is apparent in unreasonably small $\chi^{2}$ values for a variety of shock parameters, line fractions, and beam filling factors.
In addition, the beam filling factor and the line fractions are degenerate.

To obtain useful model fits it is necessary to constrain the fraction of the line emission from shocks.
This can be achieved by introducing and simultaneously fit a second gas component for the remaining line emission. 
An obvious choice for this component are PDRs, which have been traced in NGC~891 through the far-infrared line emission on large scales.
In our fitting routine for the PDR phase we include our [C~{\small I}] 370~$\mu$m line and the infrared continuum.

To scale the far-IR continuum to our $11''$ beam we use the observed morphology of the submillimeter continuum.
\citet{Serabyn.Lis.Dowell.1999} obtained a map of NGC~891 in the 350~$\mu$m continuum using SHARC on the CSO.
They estimated that about 4.8\% of the continuum flux density from the entire galaxy emerges from within the central $12''$. 
Here we assume that the total IR luminosity of NGC~891 of $L(8-1000\mu{\rm m}) = 1.89 \times 10^{10} L_{\odot}$ \citep[RBGS;][; converted to a distance of 9.5~Mpc]{Sanders.Mazzarella.Kim.2003} scales the same. 
We convert the IR luminosity from the central region into an intensity and multiply that value by 1.09 to scale to our $11''$ FWHM beam. 
A large fraction of the IR continuum emission in NGC~891 likely arises from a diffuse dust component and is probably not associated directly with PDRs.
Considering the entire galaxy as much as 69\% of the IR continuum might arise from this component \citep{Popescu.Tuffs.Kylafis.2004}.
Given that our observation is focused on the center of NGC~891 the percentage might be lower.
Here we follow \citetalias{Stacey.Charmandaris.Boulanger.2010} and assume a value of 50\% as the IR-fraction arising from cirrus in NGC~891, and allow an uncertainty of 50\%.

We exclude the [C~{\small I}] 609~$\mu$m observation (Table~\ref{tab:litobs}) since the intensity appears unreasonably small compared to the [C~{\small I}] 370~$\mu$m line intensity.
The ratio of the (converted to $11''$) [C~{\small I}] 370/[C~{\small I}] 609 intensity is 17.3.
In a PDR this ratio increases with density and FUV field, and for a density of $n = 10^{6}$ cm$^{-3}$ and a FUV field of $G_{o}=10^{6}$ (in units of the Habing field: $1.6\times10^{3}$ erg/s/cm$^{2}$) this line ratio is only 7.8 \citep{Kaufman.Wolfire.Hollenbach.1999}.
For a PDR density of $n = 10^{3}$ cm$^{-3}$ and a FUV field of $G_{o}=10^{2}$ the expected line ratio would be 2.4.
The CO(3-2) intensity quoted by \citet{Bayet.Gerin.Phillips.2006} is also the smallest value compared with the values reported by \citet{Mauersberger.Henkel.Walsh.1999} and \citet{Dumke.Nieten.Thuma.2001}.
It could be possible that our ZEUS [C~{\small I}] 370~$\mu$m intensity is too high.
However, we have measured the [C~{\small I}] 370~$\mu$m line simultaneously with the CO(7-6) line. 
If the [C~{\small I}] 370~$\mu$m intensity is too high would also mean the CO(7-6) is too high.
Both the [C~{\small I}] lines and the cirrus-subtracted IR continuum are expected to originate in PDRs.
The [C~{\small I}] 370/IR and [C~{\small I}] 609/IR intensity ratios is $2.7\times10^{-4}$ and $1.6\times10^{-5}$, respectively.
For a PDR density of $n = 10^{3}$ cm$^{-3}$ and a FUV field of $G_{o}=10^{2}$ the expected ratios are $3.2\times10^{-4}$ and $1.3\times10^{-4}$ for the [C~{\small I}] 370/IR and [C~{\small I}] 609/IR intensity ratios, respectively.
Although we can not rule out errors in estimating the IR continuum arising from PDRs in the central region of NGC~891 or a systematic error in our calibration, it appears that the ZEUS [C~{\small I}] 370~$\mu$m intensity is consistent with other measurement and PDR predictions.
A possibility for the smaller than expected [C~{\small I}] 609~$\mu$m intensity is self-absorption by cold foreground gas.
The [C~{\small I}] 609~$\mu$m transition connects to the ground, has an upper level energy equivalent to 24~K, and atomic carbon is the fourth most abundant element.
Thus, in an edge-on galaxy self-absorption of fine-structure line transitions between the lowest energy levels of abundant elements is a plausible scenario.
For example, [O~{\small I}] 63~$\mu$m line also connects to the ground, and self-absorption by colder foreground gas has also been invoked to explain the relatively weak [O~{\small I}] 63~$\mu$m emission from galaxies \citep[e.g.][and references therein]{Vasta.Barlow.Viti.2010}.

We compare the observed to the predicted line intensities from the combination of a PDR and shock model and apply a least squares algorithm to fit the individual line fractions and the beam filling factors.
The model grid for shocks \citep{Flower.PineauDesForets.2010} is rather coarse, with just two values for the pre-shock densities ($2\times10^{4}$ cm$^{-3}$ and $2\times10^{5}$ cm$^{-3}$) and velocity ranges from 10-30~km/s for J-shocks and 10-40~km/s for C-shocks, with a resolution of 10~km/s.
The model grid for the PDR parameters covers a large range, with densities, $\log(n/{\rm cm^{-3}})$, and far-UV fields, $\log(G_o)$, between 1-5, with a step size of 0.25.
In the fitting algorithm, we consider the model intensities from a single PDR surface.
We also take into account that the CO emission is usually optically thick \citep[e.g.][]{Bradford.Stacey.Nikola.2005} and that 50\% of the infrared continuum originates from cirrus in NGC~891.
For an externally FUV-irradiated molecular cloud PDRs are created on the outer surface.
Thus, along the line of sight a beam intercepts two projected PDR surfaces from a single cloud, one at the front of the cloud and one at the back of the cloud.
An optically thin line would be observed from both PDRs, while an optically thick line would only be detected from the front side.
To compare the predicted line intensities from a single PDR surface with the observations we therefore multiply the predicted intensities of the optically thin lines by a factor of 2.
The beam filling factor derived for PDRs is then with respect to single PDR surfaces and not for entire clouds.
We apply this fitting routine to two combinations: a PDR and C-type shock combination and a PDR and a J-type shock combination.
Note that the beam filling factors can be larger than unity if the beam intersects many PDR surfaces or shocks along the line of sight, with each contributing to the observed intensities.
We do not include the far-infrared ISO lines in the fitting algorithm because of the uncertain scaling factors.
Instead, we just list the predicted intensities for the [O~{\small I}], [C~{\small II}], and [C~{\small I}] 609~$\mu$m lines from the shock and PDR models.
The results are shown in Tables~\ref{tab:fit_cpdr}, \ref{tab:fit_jpdr}, and \ref{tab:fit_res}, where the predicted intensities from shocks and PDRs are already multiplied by the appropriate beam filling factor and for the PDRs, the optically thin lines are multiplied by an additional factor of 2.

The fit of combined PDR and shock models to the data is slightly better for J-type than for C-type shocks.
However, most of the resulting model parameters are similar and the difference in $\chi^2$ is not enough to be conclusive.
In both cases the best fit is for a pre-shock gas density of $n = 2 \times 10^{4}$ cm$^{-3}$, a PDR density of $n_{\rm PDR} = 3.2\times10^{3}$ cm$^{-3}$, and a FUV field of $G_{o}=100$.
For the J-shock/PDR combination, the remaining parameters that best fit the data are a shock velocity of 20~km/s, a shock beam filling factor of 0.87, and a PDR filling factor of 1.34.
For the C-shock/PDR combination the best shock velocity is 10~km/s and the beam filling factors are 0.25 for the C-shock and 1.28 for the PDR component.
The fraction of the individual line emissions is also very similar for both types of shocks.
In both cases 100\% of the CO(7-6) and 99\% of the CO(6-5) intensity would arise from shock-excited gas.
Of the CO(3-2) emission, 47\% and 32\% would arise from shocks in the C-shock/PDR model and J-shock/PDR model, respectively. 
For the H$_{2}$ emission, 94\% and 93\% of the S(2) intensity and 45\% and 44\% of the S(1) intensity would arise from shocks in the C-shock/PDR and J-shock/PDR model, respectively.
For both model combinations only about 3\% of the H$_{2}$ S(0) intensity would originate in shocks.
The main difference between the results of the model combination is in the beam filling factor of the shocks.

The parameter solution for the shock models is very sensitive to a change in observed intensities, and more observations would be very helpful to provide better constraints.
This is mainly due to the coarse grid of the shock model.
Marginalization of the probability density distribution shows that the PDR solutions are strongly peaked with a 1~sigma range less than a step-size in the model grid for density and far-UV parameters.
While the marginalized probability density distribution for the shock velocities also strongly peak at the best fit values with a 1~sigma range less than a step size, running the fitting routine with the CO(3-2) intensity increased by about 10\% changes the densities of the C-shock/PDR model to a pre-shock density of $n = 2 \times 10^{5}$ cm$^{-3}$, a shock filling factor of 0.11, and a PDR density of $n_{\rm PDR} = 5.6\times10^{3}$ cm$^{-3}$.
In contrast, J-shock/PDR solutions are hardly affected.
The fractions of the line emission originating from shocks are quite robust.
Given the uncertainty in the estimate of the IR continuum and the observed CO(3-2) intensity, for which we assumed a relative error of 50\% in the fitting routine, the predicted intensities are not better than a factor of 2 in most cases.
In fact, a change in shock velocity by 10~km/s can especially change the predicted [O~{\small I}] intensities originating from shocks by an order of magnitude.
Nevertheless, except for high-velocity J-shocks the [O~{\small I}] intensities from shocks would always be smaller compared to the predicted PDR intensities.

We note that the observed [C~{\small I}] 609~$\mu$m intensity is too weak by roughly an order of magnitude compared to the J-shock/PDR  model result and by about a factor of about three compared to the C-shock/PDR model result.

Since there are no high spatial resolution observations available for the far-IR lines we can only compare the model predictions among each other.
Including the beam filling factors, the models predict a fraction of $< 1$\% of the [O~{\small I}] 63~$\mu$m and 146~$\mu$m intensity from C-shocks or J-shocks.
For our best fit shock/PDR model combinations, the entire [O~{\small I}] emission is predicted to originate from PDRs.
It is nevertheless interesting to notice that the predicted intensity of the [O~{\small I}] 63~$\mu$m line is similar to the intensity in the H$_{2}$ S(0) line in shocks.
However, both intensities are small compared to the intensity in the higher H$_{2}$ transitions.

We now compare the predicted far-IR line emission that likely emerges from the central $11''$ with the ISO observations.
According to the CO(1-0) cut along the major axis of NGC~891 \citep[e.g.][]{Sofue.Nakai.1993}, about 20\% of the line emission within the $75''$ (FWHM) central region originates from the central $11''$ (FWHM).
If the ISO far-IR fine-structure lines scale similarly, as would be a reasonable assumption for the [C~{\small II}] line, then the predicted line intensities from our PDR/shock model are only about a factor of 2 larger than the scaled ISO far-IR lines.

\begin{figure}[ht]
\centering
\includegraphics[width=0.90\textwidth]{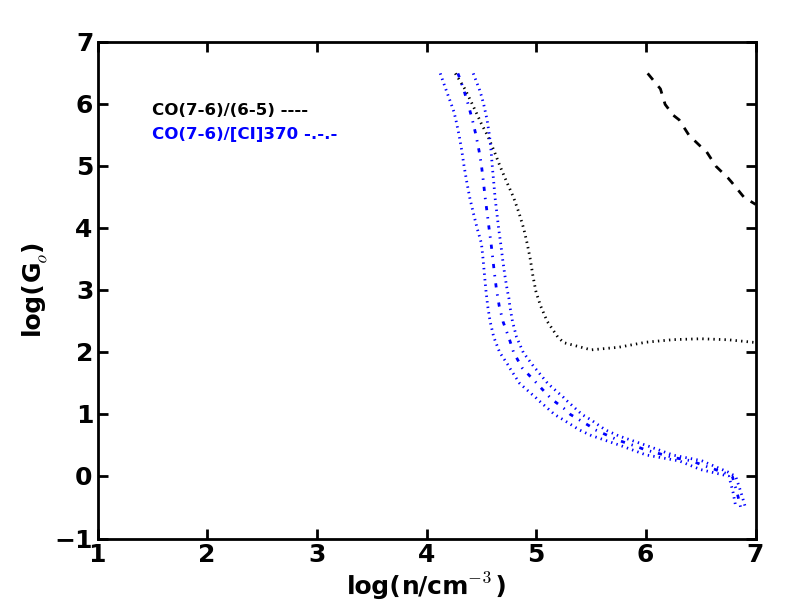}
\caption{Plot of parameter range of PDR model (FUV field, $G_{o}$, in units of Habing field $1.6\times10^{3}$ erg/s/cm$^{2}$ versus density) depicting the observed CO(7-6)/CO(6-5) (dashed) and CO(7-6)/[C~{\small I}] 370~$\mu$m (dashed-dotted) intensity ratios. The uncertainties are indicated by dotted lines.}
\label{fig:pdr}
\end{figure}

We can also just fit the PDR model grid to the ISO observations of the center of NGC~891.
If the IR continuum would scale similar to the [C~{\small II}] emission as observed by ISO then about 23\% of the IR continuum from the entire galaxy would arise from the inner $75''$ (FWHM).
As before, we also take into account that probably only 50\% of the IR continuum might arise from PDRs (with the rest originating from cirrus) and that the [O~{\small I}] 63~$\mu$m line is optically thick.
The best PDR fit to the observed [C~{\small II}], and [O~{\small I}] 63~$\mu$m and 146~$\mu$m ISO observations and the IR continuum is obtained for a gas density of $n_{\rm PDR} = 1 \times 10^{3}$ cm$^{-3}$, a FUV field of $\log(G_{o}) = 2.2$, and a PDR beam filling factor of 0.07.
The observed intensities and the intensities predicted by the PDR model are shown in Table~\ref{tab:isopdrfit}.
Both, the estimated density and far-UV field are similar to the predictions we obtained for the combined shock/PDR model within the much smaller ZEUS beam.
In contrast, the beam filling factor is much smaller than the one we obtained previously for the ZEUS observation, as would be expected for the large ISO beam, whose diameter is much larger than the thickness of the edge-on galactic disc.
This small PDR beam filling factor within the ISO beam would correspond to a filling factor of 3.26 within the ZEUS beam.
It thus appears that the PDR conditions are similar within a factor of a few ($<10$) within small region covered by ZEUS and the large region covered by ISO.
If we apply the beam filling factor corresponding to the ZEUS beam (3.26) instead of the ISO beam filling factor, which translates into multiplying the predicted values in Table~\ref{tab:isopdrfit} by $3.26/0.07$,  then the predicted intensities from the PDR are within factors of 2-3 for the [C~{\small I}] 370~$\mu$m intensity observed with ZEUS and the CO(3-2) and H$_{2}$ S(0) intensities.
The PDR model, however, fails to predict the mid-J CO intensities by more than 2 orders of magnitude and the H$_{2}$ S(2) emission by about 1 order of magnitude.
In fact, the large observed CO(7-6)/CO(6-5) intensity ratio can hardly be modeled by a PDR at all.
There is no single PDR model that can fit both the CO(7-6)/CO(6-5) and CO(7-6)/[C~{\small I}] 370~$\mu$m intensity ratio (Fig.~\ref{fig:pdr}), clearly indicating the presence of an additional heating mechanism.

\section{Energy Sources \label{sec:esource}}

We have shown that the warm interstellar medium in the nuclear region of NGC~891 is likely heated by a combination of FUV radiation and shocks, with the mid-J CO transitions probably entirely due to shocks.
Since NGC~891 is an isolated galaxy with only a moderate star formation rate (SFR) and no active galactic nucleus we do not consider shock excitation through powerful mechanism that are encountered in ULIRGs or galaxy interactions, where strong X-ray emission from an active nucleus, nuclear jets, or galaxy-wide gas flows likely drive internal or external shock fronts that are suggested by their strong H$_{2}$ emission.
Instead, we have considered moderate internal energy sources as heating mechanisms for the PDRs, shocks, and turbulence.

NGC~891 has a modest SFR of 3.8~$M_{\odot}$/yr \citep{Popescu.Tuffs.Kylafis.2004}, which could be responsible for the PDR excitation of the far-IR and submillimeter fine-structure lines. 
Sources of shocks and turbulence that could excite the H$_{2}$ and CO rotational line emission include cloud-cloud collisions, outflows from young stellar objects, or supernovae driven outflows.
For example, the warm molecular gas that strongly emits mid-J CO lines in the circumnuclear ring in the Milky Way is likely heated by dissipation of MHD turbulence \citep{Bradford.Stacey.Nikola.2005}.
Cloud-cloud collisions in the material that falls into the central region is possibly the source for the turbulence there. 
However, a 3~pc scale circumnuclear ring, like the one in the Milky Way, would not contribute significantly within our 500~pc scale ZEUS beam on NGC~891.

In the center of NGC~253 the mid-J CO emission is probably either due to the decay of turbulent shocks or due to cosmic ray heating \citep{HaileyDunsheath.Nikola.Stacey.2008}.
Both mechanisms could provide enough energy to heat the warm molecular gas in NGC~253.
The enhanced cosmic ray flux in the center of NGC~253 is likely due to the high supernova rate (SNRa) of 0.1~yr$^{-1}$ in the central $\sim100$~pc \citep{Bradford.Nikola.Stacey.2003}. 
Supernova outflows could also drive turbulence.

\citet{Roussel.Helou.Hollenbach.2007} have studied H$_{2}$ emission in the SINGS sample, which is comprised mainly of ``normal'' galaxies and some LINERs/Seyferts. 
For the galaxies powered by star formation they found a good correlation between the combined power in the H$_{2}$ S(0), S(1), and S(2) lines and the strength of the 7.7~$\mu$m PAH emission.
In contrast, the H$_{2}$/PAH ratio in the LINER/Seyfert galaxies shows a much larger scatter and the ratio is usually elevated for stronger H$_{2}$ emission.
This suggests that if the combined power in the H$_{2}$ S(0), S(1), and S(2) lines is above a certain threshold, in relationship to the strength of the 7.7~$\mu$m PAH feature, then the H$_{2}$ lines are too bright to be excited exclusively within PDRs. 
For H$_{2}$ sources above this threshold, they suggest alternative sources of heat ranging from the X-rays from a nuclear engine to interactions with nuclear outflows via shocks. 
\citetalias{Stacey.Charmandaris.Boulanger.2010} have compared the combined H$_{2}$ rotational lines with the 7.7~$\mu$m PAH feature along the plane of NGC~891 and found that the H$_{2}$/PAH ratios are near the top (threshold) values for star forming galaxies in the SINGS sample for regions within $\sim5$~kpc of the nucleus.  
These inner regions contain most ($>80$\%) of the observed H$_{2}$ line luminosity for the galaxy so that most of the line emission is at the threshold between PDR and shock excited origins. 
However, for the outer regions of NGC~891 the H$_{2}$/PAH ratio appears to rise by a factor of order 2 which places these regions more into the shock excited regime as defined by \citet{Roussel.Helou.Hollenbach.2007}. 
The rise in the H$_{2}$/PAH ratio is due to a faster decline in the PAH emission compared with the H$_{2}$ emission in the outer galaxy, and might be due to a variety of effects including a lessened importance of star formation in the outer galaxy, or lowered metallicity in the outer regions of NGC~891.  
This later effect is a plausible cause for the enhanced H$_{2}$/PAH ratio seen for the low metallicity dwarf galaxy NGC~6822 in the \citet{Roussel.Helou.Hollenbach.2007} sample. 

For our modeled C-type shocks with a velocity of 10~km/s and a pre-shock gas density of $2\times10^{4}$ cm$^{-3}$ the total expected CO intensity (up to $J=20$) is $3.2\times10^{-4}$ erg/s/cm$^{2}$/sr \citep[Table A1 in][]{Flower.PineauDesForets.2010}.
The total intensity of the H$_{2}$ rotational transitions is a factor of 2 higher, and together with CO they provide basically the entire radiative cooling for the C-shock.
Flux energy in the neutral oxygen line is less than 1\% and therefore negligible.
About 54\% of the mechanical flow energy is converted into line radiation.
Applying the derived beam filling factor of 0.25 for C-shocks, the total combined luminosity of H$_{2}$ and CO within the $11''$ beam is then $2.17\times10^{6} \ L_{\odot}$, which is roughly half the mechanical heating rate.

For our J-shock solution the total expected CO intensity (up to $J=20$) is $1.4\times10^{-3}$ erg/s/cm$^{2}$/sr \citep[Table A2 in][]{Flower.PineauDesForets.2010}, a factor of 4.3 higher than for the C-shocks.
The strongest CO transitions are around $J_{\rm up}=16$, so the total CO intensity given above is a lower limit.
The energy in all H$_{2}$ lines is a factor of about 8 higher than from all CO lines \citep{Flower.PineauDesForets.2010}.
This results in a combined luminosity of all the H$_{2}$ and CO lines of $9.7\times10^{7} \ L_{\odot}$ within the $11''$ beam and including the shock filling factor of 0.87.
Both lines account for about 82\% of the cooling, and the combined luminosity is more than 40 times larger than for C-shocks.
H$_{2}$O would account for an additional 15\% cooling, and about 97\% of the flow energy of this J-shock is converted into line radiation.
As for the C-type shocks cooling via the neutral oxygen lines is negligible for J-shocks with the above parameters.

In NGC~891, the total SNRa is about 0.042~yr$^{-1}$ \citep{Strickland.Heckman.Colbert.2004}.
Following \citet{Loenen.Spaans.Baan.2008} and assuming an energy output of $10^{51}$~erg per supernova and a transfer efficiency of 10\% the expected mechanical heating rate is $L_{\rm SN} = 8.34\times10^{9} \times 0.1 \times SNRa \ L_{\odot}$ \citep[eq.4 in][]{Loenen.Spaans.Baan.2008}, or $L_{\rm SN} = 3.5\times10^{7} \ L_{\odot}$.
This is a factor of 10 larger than the required heating rate for the C-shocks, but about a factor of 3 less than required to explain the J-shocks.

Another possibility for the heating source are outflows from young stellar objects (YSOs).
Using equation 2 of \citet{Loenen.Spaans.Baan.2008}, the ejected energy rate from YSOs can be estimated as $\frac{{\rm d}E_{\rm flow}}{{\rm d}t} \approx 3\times10^{29} r_{\rm pc}^{2} v_{\rm flow}^{3} n$ erg/s, where $v_{\rm flow}$ is the outflow velocity in km/s, $n$ the density of the surrounding medium in cm$^{-3}$, and $r_{\rm pc}$ the size of the bubble (in parsec) that is affected by the outflow.
YSOs only affect the molecular ISM locally, so for a cloud with a radius of 0.1~pc \citep{Loenen.Spaans.Baan.2008} the resulting luminosity is $\sim16 \ L_{\odot}$, and thus more than $10^{5}$ YSO's would be required within a $11''$ beam ($\sim$ 500~pc)  to account for the line luminosity in the (C-type) shock-excited gas. 
Given the SFR of 3.8~$M_{\odot}$/yr \citep{Popescu.Tuffs.Kylafis.2004} and a lifetime of the YSO phase of $10^{5}-10^{6}$~yr \citep{Loenen.Spaans.Baan.2008} the approximate luminosity ranges from 6.1 to $61\times10^{6} \ L_{\odot}$, or 2.4 to 24 times the line luminosity from C-type shocks.
\citet{Lada.Lombardi.Alves.2010} suggest a relationship between the star-forming rate and the number of YSO based on observations of Galactic star-forming clouds: $N(YSO) = 4 \times 10^{6} \times SFR~M_{\odot}{\rm yr^{-1}}$.
Applying this relationship to the SFR of the entire galaxy NGC~891 we would expect about $1.5\times 10^{7}$ YSO.
NGC~891 is often thought of as an edge-on analog to the Milky Way.
Withing the central $\approx 400$~pc of the Milky Way the SFR was estimated to be $\sim0.07 M_{\odot}~{\rm yr^{-1}}$ \citep{An.Ramirez.Sellgren.2011}, which would correspond to about $3\times10^{5}$ YSO's.
If the energy transfer efficiency is similar to SN, YSO outflows could provide enough energy to drive C-type or J-type shock excitation.

\section{Conclusions \label{sec:concl}}

We have observed the CO(6-5), CO(7-6), and [C~{\small I}] 370~$\mu$m lines from the nuclear region of NGC~891 using our submillimeter grating spectrometer ZEUS on the CSO.
To analyze the data we have include observations of the CO(3-2) \citep{Mauersberger.Henkel.Walsh.1999} and H$_{2}$ S(0), S(1), S(2) \citepalias{Stacey.Charmandaris.Boulanger.2010}, and IR continuum from the literature.
We find the emission is best explained by a combination of PDRs and shocks.
While J-shocks provide a slightly better fit than C-shocks the type of shocks is not conclusive.

The CO(7-6)/CO(6-5) intensity ratios are large for a PDR paradigm and would require extremely high far-UV and density values.
In addition the CO ratio and the CO(7-6)/[C~{\small I}] 370~$\mu$m ratio can not be fitted by a single PDR model. However, the mid-J CO ratios are consistent with shock-excited warm molecular gas, and our ZEUS observations together with Spitzer H$_{2}$ observations can be explained by a combined shock/PDR model.
This result confirms recent studies of the H$_{2}$~S(0), S(1), and S(2) transitions with Spitzer \citepalias{Stacey.Charmandaris.Boulanger.2010}, which required C-shocks to explain the higher H$_{2}$ rotational transitions.
The best fitting shock models require a a pre-shock gas density of $2\times10^{4}$~cm$^{-3}$ and shock velocities of 10~km/s and 20~km/s and shock filling factors of 0.25 and 0.87 for C-shocks and J-shocks, respectively.
The most likely energy sources for the shocks in the center of NGC~891 are supernovae and outflows from YSOs.
While both types of shocks fit our data about equally well, from energy considerations it is significantly easier to provide the requisite mechanical energy for C-shock excitation of the lines than it is for J-shock excitation.

We have estimated the fraction of line emission originating from shocks and PDRs and find that for both types of shocks all of the CO(7-6) and 99\% of the CO(6-5) intensity arises from shock-excited gas.
For the CO(3-2) emission, about 47\% and 32\% would originate from C-shocks or J-shocks, respectively.
Our model fits also suggest a larger fraction of the H$_{2}$ S(2) (94\%) and S(1) (45\%) emission from shocks than estimated by \citetalias{Stacey.Charmandaris.Boulanger.2010}.
Only about 3\% of the H$_{2}$ S(0) emission would emerge from shock-excited gas.

The [C~{\small I}] 370~$\mu$m emission arises from PDRs.
When fitting the combined PDR/shock model to the ZEUS and Spitzer data we find a best solution for a PDR density of $3.2\times10^{3}$~cm$^{-3}$ and a FUV field of $G_{o} = 100$ for both types of shocks.
The PDR filling factors are 1.28 and 1.34 for a combination with C-shocks and J-shocks, respectively.
Our best PDR solution is consistent with [O~{\small I}] 63~$\mu$m and 146~$\mu$m and [C~{\small II}] observations on the scale of the larger ISO beam.
Comparing the line predictions from our best fitting shock and PDR models shows that the [O~{\small I}] emission arises almost entirely from PDRs.

\acknowledgments
We thank the CSO staff for their excellent support of the ZEUS observations. 
We also thank the anonymous referee whose comments led to a great improvement in the data analysis.
This work was supported by NSF grants AST-0096881, AST-0352855, AST-0705256, and AST-0722220, and by NASA grants NGT5-50470 and NNG05GK70H.

\bibliographystyle{apj}
\bibliography{n891_zeus}{}

\clearpage

\begin{deluxetable}{l c c c c}
\tabletypesize{\footnotesize}
\tablecaption{ZEUS Observations \label{tab:zeusobs}}
\tablehead{
\colhead{Line}  &  \colhead{Beam}  &  \colhead{$\int T_{MB} {\rm d}v$} & \colhead{$I$}  &  \colhead{$F$}  \\
\colhead{}  &  \colhead{[arcsec]}  &  \colhead{[K km/s]}  &  \colhead{[erg/s/cm$^2$/sr]}  &  \colhead{[W/m$^2$]}  
}
\startdata
CO(6-5)  &  $11''$  &  $31.2 \pm 4.0$  &  1.056E-5  &  3.403E-17  \\
CO(7-6)  &  $10''$  &  $35.4 \pm 7.8$  &  1.902E-5  &  5.066E-17  \\
$[$CI$]$ 370$\mu$m  &  $10''$  &  $65.6 \pm 7.8$  &  3.563E-5  &  9.490E-17  \\
\enddata
\end{deluxetable}

\vfill

\clearpage

\begin{deluxetable}{l c c c c c}
\tabletypesize{\footnotesize}
\tablecaption{Observations taken from the literature \label{tab:litobs}}
\tablehead{
\colhead{Line}  &  \colhead{Observation}  &  \colhead{Obs.Area}  &  \colhead{$\int T_{MB} {\rm d}v$}  &  \colhead{$I$}  &  \colhead{$F$}  \\
\colhead{}  &  \colhead{}  &  \colhead{[arcsec]}  &  \colhead{[K km/s]}  &  \colhead{[erg/s/cm$^2$/sr]}  &  \colhead{[W/m$^2$]}  
}
\startdata
H$_{2}$~S(0) \tablenotemark{a}  &  Spitzer  &  $22.3''\times11.1''$  &    &  3.39E-5  &  1.97E-16  \\
H$_{2}$~S(1) \tablenotemark{a}  &  Spitzer  &  $4.7''\times11.3''$  &    &  9.11E-5  &  1.14E-16  \\
H$_{2}$~S(2) \tablenotemark{a}  &  Spitzer  &  $4.7''\times11.3''$  &    &  4.71E-5  &  5.88E-17  \\
$[$CI$]$ 609$\mu$m \tablenotemark{b}  &  CSO  &  $14.55''$  &  $11.5 \pm 1.0$  &  1.4E-6  &  7.9E-18  \\
CO(3-2) \tablenotemark{c}  &  HHT  &  $21''$  &  $24 \pm 2$  &  1.017E-6  &  1.194E-17  \\
\enddata
\tablenotetext{a}{Extinction corrected H$_{2}$ transitions from \citetalias{Stacey.Charmandaris.Boulanger.2010}}
\tablenotetext{b}{From Table~B.1 in \citet{Bayet.Gerin.Phillips.2006}}
\tablenotetext{c}{From Table~1 in \citet{Mauersberger.Henkel.Walsh.1999}}
\end{deluxetable}

\vfill

\clearpage

\begin{deluxetable}{c c c c c c c}
\tablecaption{Results of the combined C-shock and PDR model fits \label{tab:fit_cpdr}}
\tablehead{
\multicolumn{3}{c}{C-Shock}  &  \multicolumn{3}{c}{C-PDR}  &  \\
\colhead{$\phi_{\rm C-sh}$\tablenotemark{a}}  &  \colhead{$n$ (cm$^{-3}$)}  &  \colhead{$v$ (km/s)}  &  \colhead{$\phi_{\rm PDR}$\tablenotemark{a}}  &  \colhead{$n$ (cm$^{-3}$)}  &  \colhead{FUV ($G_{o}$)}  &  \colhead{$\chi^2$}
}
\startdata
0.25  &  2.0E+4  &  10.0  &  1.28  &  3.16E+3 &  100  &  7.5  \\
\enddata
\tablenotetext{a}{$\phi_{\rm C-sh}$ and $\phi_{\rm PDR}$ are the beam filling factor for the C-shock and the PDR component, respectively.}
\end{deluxetable}

\vfill

\clearpage

\begin{deluxetable}{c c c c c c c}
\tablecaption{Results of the combined J-shock and PDR model fits \label{tab:fit_jpdr}}
\tablehead{
\multicolumn{3}{c}{J-Shock}  &  \multicolumn{3}{c}{J-PDR}  &  \\
\colhead{$\phi_{\rm J-sh}$\tablenotemark{a}}  &  \colhead{$n$ (cm$^{-3}$)}  &  \colhead{$v$ (km/s)}  &  \colhead{$\phi_{\rm PDR}$\tablenotemark{a}}  &  \colhead{$n$ (cm$^{-3}$)}  &  \colhead{FUV ($G_{o}$)}  &  \colhead{$\chi^2$}
}
\startdata
0.87  &  2.0E+4  &  20.0  &  1.34  &  3.16E+3  &  100  &  5.0  \\
\enddata
\tablenotetext{a}{$\phi_{\rm J-sh}$ and $\phi_{\rm PDR}$ are the beam filling factor for the J-shock and the PDR component, respectively.}
\end{deluxetable}

\vfill

\clearpage

\begin{deluxetable}{l c c c c c c c}
\tablecaption{Observed/converted Intensities within $11''$ beam, fraction of line emission from shocks and predicted intensities of the C-shock/PDR and J-shock/PDR model fits \tablenotemark{a} \label{tab:fit_res}}
\tablehead{
\colhead{Line}  &  \colhead{$I_{\rm obs}$}  & \colhead{f(C)\tablenotemark{b}}  &  \colhead{C-Shock}  &  \colhead{C-PDR}  &  \colhead{f(J)\tablenotemark{b}}  &  \colhead{J-Shock}  &  \colhead{J-PDR}
}
\startdata
CO(7-6)                     &  1.74E-5  &  1.00  &  1.279E-5  &  1.508E-8  &  1.00  &  1.693E-5  &  1.580E-8  \\
CO(6-5)                     &  1.06E-5  &  0.99  &  1.091E-5  &  1.016E-7  &  0.99  &  9.773E-6  &  1.064E-7  \\
CO(3-2)                     &  2.24E-6  &  0.47  &  1.469E-6  &  1.663E-6  &  0.32  &  7.774E-7  &  1.743E-6  \\
H$_{2}$ S(2)                &  4.72E-5  &  0.94  &  3.221E-5  &  2.583E-6  &  0.93  &  3.411E-5  &  2.707E-6  \\
H$_{2}$ S(1)                &  8.07E-5  &  0.45  &  3.964E-5  &  4.987E-5  &  0.44  &  3.935E-5  &  5.226E-5  \\
H$_{2}$ S(0)                &  4.17E-5  &  0.03  &  1.338E-6  &  2.299E-5  &  0.03  &  1.224E-6  &  2.410E-5  \\
$[$OI$]$63$\mu$m            &           &        &  2.725E-6  &  2.809E-4  &        &  1.399E-6  &  2.943E-4  \\
$[$OI$]$146$\mu$m           &           &        &  2.477E-7  &  2.788E-5  &        &  3.935E-8  &  2.922E-5  \\
$[$CI$]$370$\mu$m           &  3.27E-5  &        &            &  3.138E-5  &        &            &  3.298E-5  \\
$[$CI$]$609$\mu$m           &  1.89E-6  &        &            &  9.647E-6  &        &            &  1.011E-5  \\
Total IR\tablenotemark{c}   &  1.21E-1  &        &            &  6.650E-2  &        &            &  6.969E-2  \\
$[$CII$]$  &           &        &            &  6.106E-4  &        &            &  6.399E-4  \\
\enddata
\tablenotetext{a}{
To compare model values with observed values we assumed single PDR surfaces in the model fits. The model values of optically thin lines are multiplied by a factor of 2, which assumes that optically thin emission arises from twice as many PDR surfaces as optically thick emission. Optically thin lines are: all H$_{2}$ lines, [O~{\small I}] 146$\mu$m, both [C~{\small I}] lines and IR continuum. The predicted model intensities given in the table are also multiplied by the corresponding beam filling factor. Intensities are in units: erg/s/cm$^{2}$/sr.}
\tablenotetext{b}{Fraction of observed intensity arising from either C-shock or J-shock. Note that the shock and PDR intensities given are values from a model grid with limited resolution and thus don't have to mirror the exact intensity fractions for the observed intensities as derived by the fitting algorithm.}
\tablenotetext{c}{IR (8-1000$\mu$m) continuum scaled according to the 350~$\mu$m continuum SHARC map \citet{Serabyn.Lis.Dowell.1999}; see text. About 50\% of the infrared continuum likely originates from cirrus clouds in NGC~891. The value of the observed/converted IR intensity is not corrected for the cirrus emission, yet. However, model intensities were fitted to the corrected IR intensity.}
\end{deluxetable}

\vfill

\clearpage

\begin{deluxetable}{l c c}
\tablecaption{Observed intensities compared with predicted PDR intensities within ISO beam \tablenotemark{a} \label{tab:isopdrfit}}
\tablehead{
\colhead{Line}  &  \colhead{Observed}  &  \colhead{PDR}
}
\startdata
$[$CII$]$ \tablenotemark{b}  &  3.586E-5  &  3.639E-5  \\
$[$OI$]$ 63$\mu$m  &  1.377E-5  &  1.437E-5  \\
$[$OI$]$ 146$\mu$m  &  1.802E-6  &  1.561E-6  \\
Total IR\tablenotemark{c}  &  6.266E-3  &  6.525E-3  \\
$[$CI$]$ 370$\mu$m  &    &  1.249E-6  \\
$[$CI$]$ 609$\mu$m  &    &  5.031E-7  \\
CO(7-6)  &    &  3.998E-11  \\
CO(6-5)  &    &  2.684E-10  \\
CO(3-2)  &    &  2.695E-8  \\
H$_{2}$ S(2)  &    &  8.030E-8  \\
H$_{2}$ S(1)  &    &  1.947E-6  \\
H$_{2}$ S(0)  &    &  5.511E-7  \\
\enddata
\tablenotetext{a}{To compare model values with observed values we assumed single PDR surfaces in the model fits. The model values of optically thin lines are multiplied by a factor of 2, which assumes that optically thin emission arises from twice as many PDR surfaces as optically thick emission. Optically thin lines are: all H$_{2}$ lines, [O~{\small I}] 146$\mu$m, both [C~{\small I}] lines and IR continuum. The predicted intensities are also multiplied by the beam filling factor (0.07). Intensities are in units: erg/s/cm$^{2}$/sr}
\tablenotetext{b}{The observed [C~{\small II}] intensity is multiplied by 0.7, which assumes that only 70\% of the line originates from PDRs}
\tablenotetext{c}{50\% of the infrared continuum likely arises from cirrus clouds in NGC~891, which is corrected in the model values.}
\end{deluxetable}

\vfill

\clearpage

\vfill

\clearpage

\vfill

\end{document}